\documentclass[lettersize,journal]{IEEEtran}
\usepackage{amsmath,amsfonts}
\usepackage{array}
\usepackage{subfig}
\usepackage{stfloats}
\usepackage{url}
\usepackage{verbatim}
\usepackage{graphicx}
\usepackage{cite}
\usepackage{microtype}
\usepackage{doi}
\usepackage{float}
\usepackage{mathrsfs}
\usepackage{threeparttable}
\usepackage{booktabs}
\usepackage{multicol}
\usepackage{multirow}
\usepackage{tabto}
\usepackage[ruled]{algorithm2e}
\usepackage{algpseudocode}
\usepackage{makecell}
\usepackage{array}
\usepackage{diagbox}
\usepackage{adjustbox}
\usepackage{hyperref}
\usepackage{textcomp}
\usepackage{stackrel,amssymb}
\usepackage{xcolor}

\begin{document}
 \title{MoEDiff-SR: Mixture of Experts-Guided Diffusion Model for Region-Adaptive MRI Super-Resolution}
 
\author{Zhe Wang, Yuhua Ru, Aladine Chetouani, Fang Chen, Fabian Bauer, Liping Zhang, Didier Hans$^*$, Rachid Jennane$^*$, Mohamed Jarraya$^*$, Yung Hsin Chen$^*$
\thanks{Zhe Wang, Yung Hsin Chen and Mohamed Jarraya are with Department of Radiology, Massachusetts General Hospital, Harvard Medical School, Boston, 02114, USA (e-mail: zwang78@mgh.harvard.edu; ychen4@mgh.harvard.edu; mjarraya@mgh.harvard.edu).}
\thanks{Yuhua Ru is with Jiangsu Institute of Hematology, The First Affiliated Hospital of Soochow University, Suzhou, 215006, China. (e-mail: ruyuhua@suda.edu.cn).}
\thanks{Aladine Chetouani is with L2TI Laboratory, University Sorbonne Paris Nord, Villetaneuse, 93430, France (e-mail: aladine.chetouani@univ-paris13.fr).}
\thanks{Fang Chen is with department of Medical School, Henan University of Chinese Medicine, Zhengzhou, 450046, China. (e-mail: chenfangyxy@hactcm.edu.cn).}
\thanks{Fabian Bauer is with Division of Radiology, German Cancer Research Center, Heidelberg, 69120, Germany (e-mail: fabian.bauer@dkfz-heidelberg.de).}
\thanks{Liping Zhang is with Athinoula A. Martinos Centre for Biomedical Imaging, Massachusetts General Hospital, Harvard Medical School, Boston, 02114, USA. (e-mail: lzhang90@mgh.harvard.edu).}
\thanks{Didier Hans is with Nuclear Medicine Division, Geneva University Hospital, Geneva, 1205, Switzerland. (e-mail: didier.hans@chuv.ch).}
\thanks{Rachid Jennane is with IDP Institute, UMR CNRS 7013, University of Orleans, Orleans, 45067, France (e-mail: rachid.jennane@univ-orleans.fr).}}

\markboth{Journal of \LaTeX\ Class Files,~Vol.~14, No.~8, August~2021}%
{Shell \MakeLowercase{\textit{et al.}}: A Sample Article Using IEEEtran.cls for IEEE Journals}

\maketitle
\begin{abstract}
Magnetic Resonance Imaging (MRI) at lower field strengths (e.g., 3T) suffers from limited spatial resolution, making it challenging to capture fine anatomical details essential for clinical diagnosis and neuroimaging research. To overcome this limitation, we propose MoEDiff-SR, a Mixture of Experts (MoE)-guided diffusion model for region-adaptive MRI Super-Resolution (SR). Unlike conventional diffusion-based SR models that apply a uniform denoising process across the entire image, MoEDiff-SR dynamically selects specialized denoising experts at a fine-grained token level, ensuring region-specific adaptation and enhanced SR performance. Specifically, our approach first employs a Transformer-based feature extractor to compute multi-scale patch embeddings, capturing both global structural information and local texture details. The extracted feature embeddings are then fed into an MoE gating network, which assigns adaptive weights to multiple diffusion-based denoisers, each specializing in different brain MRI characteristics, such as centrum semiovale, sulcal and gyral cortex, and grey–white matter junction. The final output is produced by aggregating the denoised results from these specialized experts according to dynamically assigned gating probabilities. Experimental results demonstrate that MoEDiff-SR outperforms existing state-of-the-art methods in terms of quantitative image quality metrics, perceptual fidelity, and computational efficiency. Difference maps from each expert further highlight their distinct specializations, confirming the effective region-specific denoising capability and the interpretability of expert contributions. Additionally, clinical evaluation validates its superior diagnostic capability in identifying subtle pathological features, emphasizing its practical relevance in clinical neuroimaging. Our code is available at \url{https://github.com/ZWang78/MoEDiff-SR}.
\end{abstract}


\section{Introduction}
\IEEEPARstart{M}{agnetic} resonance imaging (MRI) is a fundamental and versatile medical imaging modality that provides detailed and high-resolution visualization of soft tissues by leveraging the interaction between strong magnetic fields, radiofrequency pulses, and the intrinsic magnetic properties of biological tissues \cite{geethanath2019accessible}. Unlike ionizing radiation-based techniques such as Computed Tomography (CT) or X-ray imaging, MRI offers a non-invasive and radiation-free approach, making it particularly suitable for repeated examinations and longitudinal studies \cite{niraj2016mri}. Its exceptional ability to generate high-contrast images with superior tissue differentiation has positioned it as an indispensable tool in the diagnosis, monitoring, and treatment planning of a wide spectrum of neurological disorders. These include neurodegenerative diseases such as Alzheimer’s and Parkinson’s, where MRI aids in detecting brain atrophy, white matter changes, and abnormal protein aggregations; demyelinating disorders like multiple sclerosis (MS), where it facilitates the visualization of plaques and disease progression; and acute cerebrovascular conditions such as ischemic stroke, where MRI enables the timely identification of infarcts and haemorrhages. As medical imaging continues to evolve, the reliance on MRI has grown substantially, driven by its ability to capture a broad spectrum of structural, functional, and biochemical changes that are crucial for both early diagnosis and prognostic evaluations \cite{skandsen2011prognostic}.

Over the past few decades, field strengths have evolved from sub-0.5T systems to the widespread adoption of 1.5/3T scanners, and more recently, to the cutting-edge development of ultra-high-field (UHF) 7T MRI \cite{nikpanah2023low}. These improvements have significantly enhanced image clarity, enabling the detection of finer anatomical structures and subtle pathological markers that remain elusive in lower-field scans \cite{vachha2021mri}. In conditions such as epilepsy, 7T MRI has demonstrated unparalleled precision in identifying small cortical lesions, thereby refining surgical planning \cite{unparalleled}. Similarly, it has also improved the visualization of cortical demyelination for MS, which is often underrepresented at conventional field strengths \cite{demyelination}. Despite its diagnostic superiority, 7T MRI faces substantial barriers to widespread clinical implementation. The high cost of procurement, installation, and maintenance poses a significant financial challenge, limiting its availability to select research institutions and specialized medical centres \cite{jones2021neuroimaging}. Additionally, the operational complexities associated with ultra-high-field imaging, such as increased susceptibility artefacts, specific absorption rate (SAR) limitations, and the necessity for advanced RF coil technology, further hinder its routine clinical adoption \cite{hoff2019safety}.


The evolution of Super-Resolution (SR) techniques in MRI has undergone a paradigm shift with the emergence of deep learning architectures, surpassing the capabilities of conventional interpolation and reconstruction methodologies. Modern deep learning approaches employ sophisticated neural networks to establish nonlinear mappings between Low-Resolution (LR) and High-Resolution (HR) image domains, achieving unprecedented fidelity in image reconstruction from lower-field MRI acquisitions \cite{chandra2021deep}. The transition from handcrafted priors to data-driven feature extraction enables the preservation of intricate anatomical details while enhancing spatial resolution, a critical advancement for clinical diagnostics and quantitative analysis \cite{ahishakiye2021survey}. Early implementations of deep learning-based SR adopted convolutional neural networks (CNNs) \cite{cnn} as foundational architectures. The pioneering work of Dong et al. with Super-Resolution Convolutional Neural Networks (SRCNN) \cite{dong2014learning} established an end-to-end learning framework that directly transformed LR MRI inputs into HR outputs through hierarchical feature abstraction. Subsequent CNN variants, including Very Deep Super-Resolution (VDSR) \cite{kim2016accurate} and Super-Resolution Residual Networks (SRResNet) \cite{ledig2017photo}, demonstrated quantitative improvements in Peak Signal-to-Noise Ratio (PSNR) by 2-4 dB and Structural Similarity Index (SSIM) enhancements of 0.05-0.12 compared to bicubic interpolation baselines. However, these architectures exhibit intrinsic limitations in modelling long-range spatial dependencies and preserving high-frequency textures, often resulting in over-smoothed reconstructions with compromised edge definition \cite{wang2020deep}. The field has witnessed significant advancements through the integration of Generative Adversarial Networks (GANs) \cite{GAN}, addressing the perceptual quality limitations of conventional CNN architectures. GAN-based frameworks, particularly those employing perceptual loss functions and attention mechanisms, have demonstrated superior performance in texture synthesis and structural preservation. Ledig et al. \cite{ledig2017photo} demonstrated that SRGAN architectures could achieve Mean Opinion Scores (MOS) improvements of 0.8-1.2 points compared to CNN-based methods through adversarial training strategies. Furthermore, the Enhanced Super-Resolution Generative Adversarial Network (ESRGAN) \cite{wang2018esrgan} proposed by Wang et al improved upon the original SRGAN by introducing a Residual-in-Residual Dense Block (RRDB) without batch normalization, allowing for deeper network architectures. ESRGAN also employed a relativistic discriminator to predict the probability that a given image is more realistic than a reference image, rather than distinguishing between real and fake images, achieving higher Natural Image Quality Evaluator (NIQE) scores compared to previous methods. K Diffusion MRI (KDMRI) \cite{kawar2022denoising} incorporates k-space consistency through a modified reverse process that minimizes data fidelity loss at each denoising step. This approach demonstrated 23$\%$ improvement in Normalized Root Mean Square Error (NRMSE) compared to vanilla diffusion models when reconstructing 7T-like images from 3T MRI inputs, while maintaining spectral compatibility in the frequency domain. The method's alternating projection between image space denoising and k-space constraint enforcement effectively reduces Gibbs artefacts in SR reconstructions. In \cite{wang2023inversesr}, InverseSR is proposed for 3D brain MRI super-resolution that leverages a pre-trained 3D Latent Diffusion Model (LDM) \cite{latent_diffusion_model} to enhance the resolution of clinical MRI scans. By utilizing the LDM as a generative prior, InverseSR captures the prior distribution of 3D T1-weighted brain MRIs, enabling the reconstruction of high-resolution images from low-resolution inputs. Validation on over 100 brain T1-weighted MRIs from the IXI dataset demonstrated that InverseSR achieved superior performance compared to baseline models.

Despite these advances, existing diffusion-based SR models in the medical imaging domain still face several challenges. Most models apply a uniform denoising strategy across the entire image, disregarding the significant anatomical and textural heterogeneity present within different regions, such as the smooth intensity transitions observed in white matter, intricate folding patterns in the cortex, and delicate branching structures within the vasculature. Such a global denoising approach inherently neglects region-specific variations in noise levels, anatomical complexity, and tissue contrast, which consequently induces excessive smoothing of fine-grained anatomical features and clinically relevant structural boundaries. Furthermore, the inability of uniform strategies to adaptively handle diverse noise characteristics often leads to persistent residual noise artefacts, adversely impacting the clinical utility of reconstructed images by potentially obscuring subtle pathological indicators or essential diagnostic markers. To address these challenges, we introduce MoEDiff-SR, a novel Mixture-of-Experts (MoE)-Guided Diffusion Model for region-adaptive MRI SR. In contrast to conventional SR methods, MoEDiff-SR dynamically assigns weights to specialized denoising experts, thereby tailoring the reconstruction process to specific anatomical structures and tissue characteristics. Specifically, MoEDiff-SR employs an MoE framework \cite{jacobs1991adaptive} wherein each diffusion expert is trained to effectively reconstruct distinct tissue types and structural patterns inherent in MRI scans. A gating network intelligently routes 3T MRI inputs to three specialized diffusion experts based on multi-scale feature embeddings derived via a Transformer-based feature extractor from the conditional input 3T MRI. These experts collaboratively generate a weighted super-resolution output, facilitating fine-grained adaptation to various anatomical regions and diverse noise levels. During the training phase, in addition to utilizing 7T MRI inputs, we incorporate gradient nonlinearity correction and bias field correction. These corrections enhance the gating network's convergence by addressing signal distortions and intensity inhomogeneities. During inference, MoEDiff-SR efficiently generates high-quality 7T-like MRI from only a 3T MRI input combined with Gaussian-distributed noise. Furthermore, the specialized structure inherent in the MoE architecture supports asynchronous inference, allowing individual experts to independently process different image regions, substantially enhancing deployability and computational efficiency. Experimental validation demonstrates that MoEDiff-SR significantly outperforms state-of-the-art SR methods, exhibiting improvements in perceptual quality metrics. Visual comparisons highlight its enhanced capacity to preserve cortical boundaries, delineate tissue interfaces, and suppress noise in homogeneous regions. Moreover, difference-map visualizations demonstrate that each expert is effectively activated in the anatomical domains for which it is specialized, validating the interpretability and targeted functionality of the MoE framework. The experimental results not only validate the technical soundness of MoEDiff-SR but also underscore its clinical relevance.

The primary contributions of this study include the following:
\begin{itemize}
\item[$\bullet$] \textbf{Token-based expert selection:} We introduce a gating mechanism that computes token-wise expert activation scores based on multi-scale patch embeddings. This enables fine-grained, region-aware expert selection, allowing the model to dynamically adapt to diverse anatomical and textural features across brain regions.
\item[$\bullet$] \textbf{MoE for adaptive diffusion denoising:} A novel MoE framework dynamically combines specialized diffusion denoisers tailored for specific anatomical characteristics, significantly enhancing the preservation of fine anatomical details.
\item[$\bullet$] \textbf{Integration of anatomical priors:} We incorporate gradient nonlinearity and bias field corrections into the training phase, significantly enhancing the gating network's ability to accurately assign experts based on precise anatomical and tissue-specific characteristics, thereby improving perceptual fidelity and structural accuracy.
\item[$\bullet$] \textbf{Clinical validation:} Through the real clinical evaluation, our model demonstrates enhanced diagnostic performance in detecting subtle pathologies that are typically obscured in lower-resolution MRI, thereby affirming its clinical utility and diagnostic relevance.
\end{itemize}

\section{Methodology}
\begin{figure}[htbp]
\centering 
\includegraphics[width=0.485\textwidth]{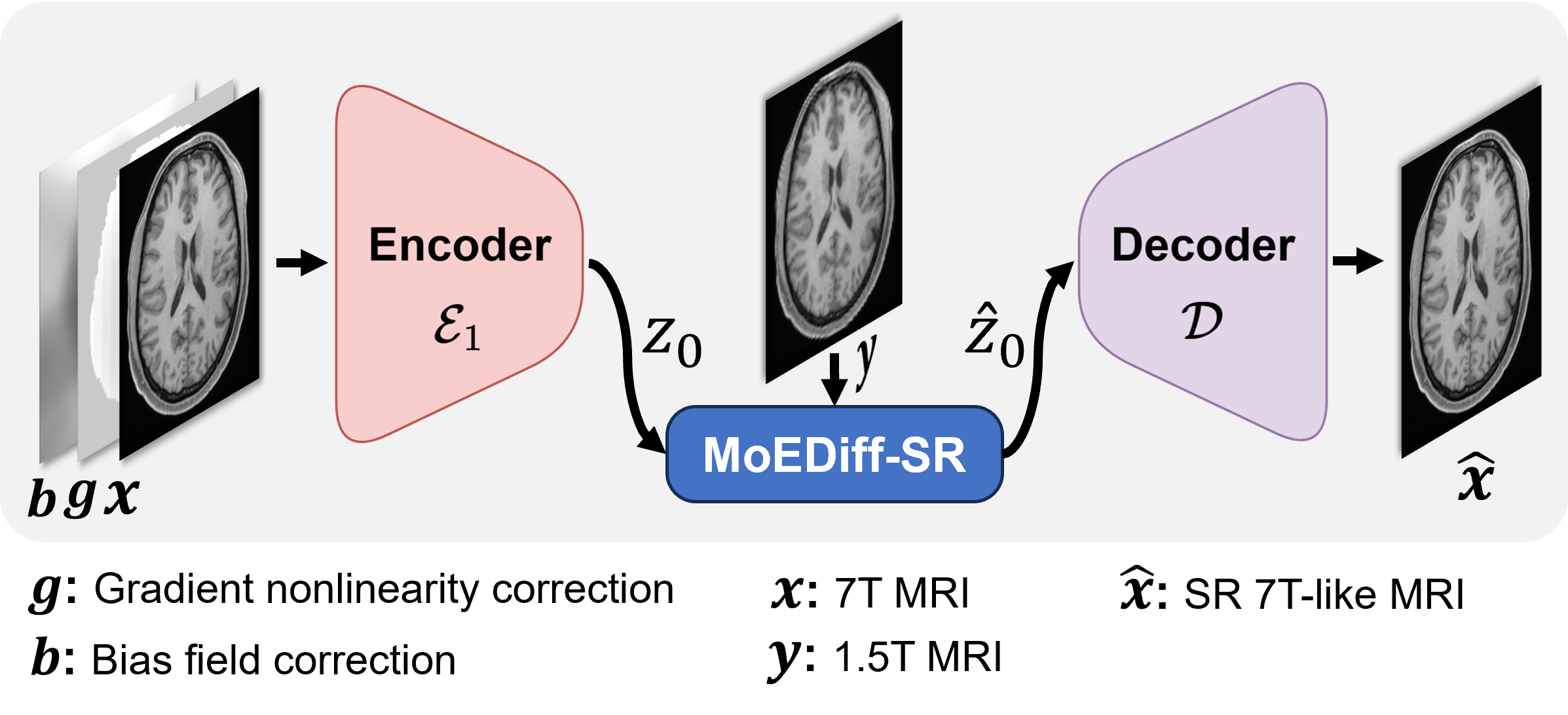}
\caption{The flowchart of the proposed global methodology.}
\label{flowchart_main}
\end{figure}

\begin{figure*}[htbp]
\centering 
\includegraphics[width=1\textwidth]{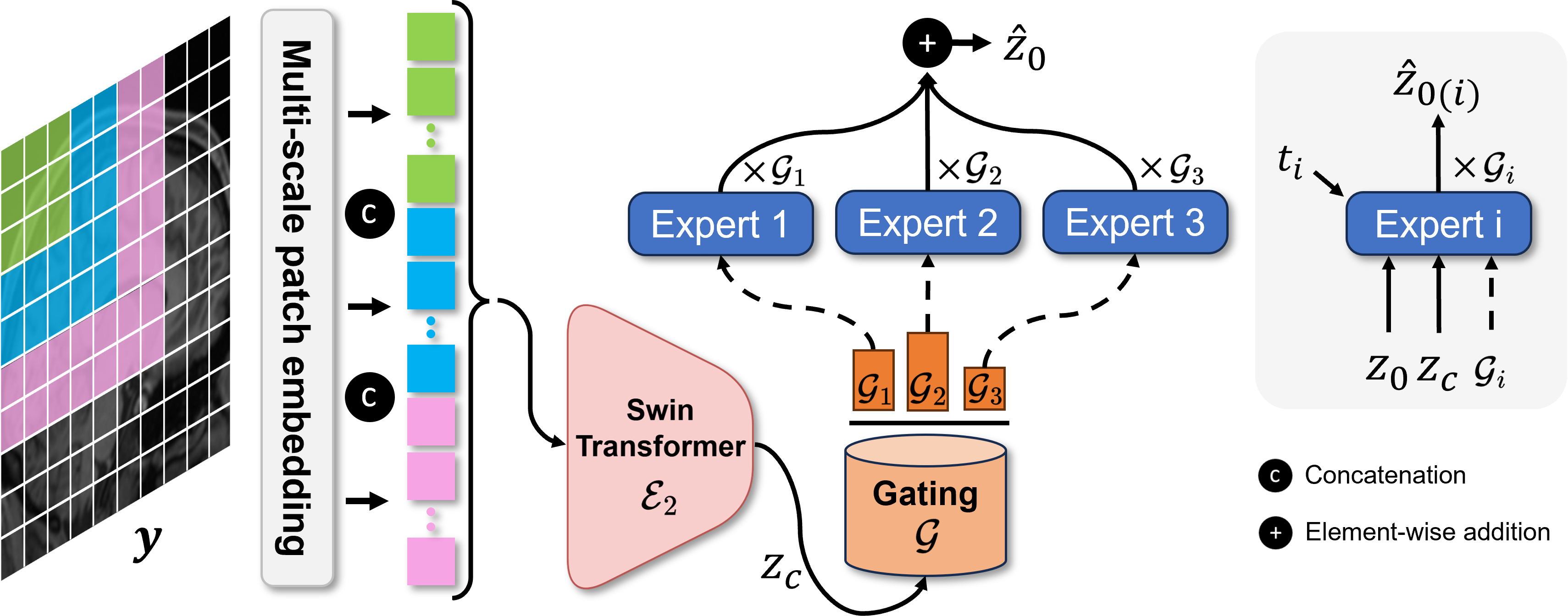}
\caption{The proposed MoEDiff-SR workflow starts by extracting multi-scale patches from the conditional input 3T MRI slice $y$. These patches are subsequently transformed via a linear projection layer, and the obtained embeddings are further encoded by a Swin Transformer $\mathcal{E}_2$, resulting in conditional latent representation $z_c$. Next, the gating network $\mathcal{G}$ dynamically computes adaptive weights ($\mathcal{G}_1$, $\mathcal{G}_2$, and $\mathcal{G}_3$) for three specialized diffusion-based SR experts ($E_1$, $E_2$, and $E_3$). Each expert follows the fundamental architecture of the Conditional Latent Diffusion Model (CLDM) \cite{latent_diffusion_model} and receives the latent representation $z_0$ from the 7T MRI, $z_c$ from the 3T MRI, the timestep $t$, and its assigned weight $\mathcal{G}_i$. The experts then individually generate region-specific denoised latent outputs $\hat{z}_{0(i)}$. Finally, these outputs are aggregated based on the dynamically assigned weights to produce the final weighted latent denoised code $\hat{z}_0$.}
\label{flowchart}
\end{figure*}

The main flowchart is illustrated in Fig. \ref{flowchart_main}. Initially, given a 7T MRI slice $x$, along with its corresponding gradient nonlinearity correction $g$ and bias field correction $b$, is processed through an image encoder $\mathcal{E}_1$ to generate the latent representation $z_0$. Subsequently, the corresponding 3T MRI slice $y$ serves as conditional input to the proposed MoEDiff-SR module, illustrated in Fig. \ref{flowchart}. Finally, the resultant latent output $\hat{z}_0$ from MoEDiff-SR is decoded by $\mathcal{D}$ to produce a super-resolved 7MRI slice $\hat{x}$ with 7T-like resolution and quality. The image encoder and decoder utilized in this study are derived from the VQ-VAE architecture \cite{van2017neural}. Given that this study primarily emphasizes the MoEDiff-SR module, its detailed description and analysis are provided in the subsequent sections.

\subsection{Multi-scale patch embedding}
\label{experts}
MRI scans exhibit complex anatomical variations across different regions. To effectively capture both global contextual information and local texture details, we employ a multi-scale patch embedding strategy followed by Transformer-based feature extraction and adaptive denoising using a MoE framework. Specifically, given a conditional input 3T MRI slice $y$, we decompose it into patches at multiple spatial scales, ensuring that the model captures features at different levels of granularity:

\begin{itemize}
    \item Large-scale patches (64×64): Encode global anatomical coherence and structural relationships.
    \item Medium-scale patches (32×32): Capture mid-level contextual features, providing a balance between spatial context and localized information.
    \item Small-scale patches (16×16): Preserve fine-grained details, critical for structures like cortical boundaries and vascular features.
\end{itemize}

Each set of patches is passed through a dedicated linear convolutional projection layer with kernel size and stride equal to the patch size, performing non-overlapping tokenization, which generates multi-scale embeddings $F_{\text{large}}, F_{\text{medium}}, F_{\text{small}}$. These multi-scale patch embeddings are concatenated along the token dimension to form a comprehensive feature representation $F_{\text{multi-scale}}$, which is then processed through the Swin Transformer encoder \cite{liu2021swin}, donated as $\mathcal{E}_2$, effectively modelling both long-range dependencies and local spatial relationships in the MRI slice:

\begin{equation}
z_c = \mathcal{E}_2(F_{\text{multi-scale}}), \quad z_c \in \mathbb{R}^{N \times d}
\end{equation}
where $z_c$ denotes the Transformer-encoded feature tokens, $N$ is the total number of non-overlapping patches aggregated from all scales, and $d$ is the feature embedding dimension of each token.

\subsection{Gating Mechanism}
\label{sec:gating_tokenwise}
To enable anatomically adaptive and semantically meaningful expert selection, we propose a token-aware gating mechanism that computes routing probabilities based on fine-grained interactions between expert-specific queries and Transformer-encoded token representations. Given the encoded token sequence $z_c \in \mathbb{R}^{N \times d}$, we first project each token through a shared lightweight MLP to obtain a refined representation $\tilde{z}_c \in \mathbb{R}^{N \times d}$, where $\tilde{z}_c^n \in \mathbb{R}^d$ denotes the $n$-th token. Then, each expert $E_i$ is associated with a learnable query vector $\mathbf{q}_i \in \mathbb{R}^d$. We compute token-wise attention weights via cosine similarity followed by softmax normalization:

\begin{equation}
Att_{n(i)} = \frac{\exp\left(\cos\theta(\mathbf{q}_i, \tilde{z}_c^n)\right)}{\sum_{m=1}^{N} \exp\left(\cos\theta(\mathbf{q}_i, \tilde{z}_c^m)\right)}
\end{equation}

The expert activation score is then calculated as the attention-weighted semantic response:

\begin{equation}
\text{Score}_i = \sum_{n=1}^{N} Att_{n(i)} \cdot \cos\theta(\mathbf{q}_i, \tilde{z}_c^n)
\end{equation}

Finally, to encourage balanced expert utilization and prevent the collapse of routing to a small subset of experts, we apply a frequency-aware softmax to compute the final routing probability for each expert:

\begin{equation}
\mathcal{G}_i = \frac{\exp(\text{Score}_i)}{\sum_j \exp(\text{Score}_j)} \cdot \frac{e^{-\gamma c_i}}{\sum_j e^{-\gamma c_j}}
\end{equation}
where $c_i$ denotes the exponentially smoothed usage frequency of expert $E_i$, which is updated during training using an Exponential Moving Average (EMA) \cite{hunter1986exponentially} with a decay factor of 0.99. $\gamma$ controls the strength of the frequency regularization (set to $\gamma = 0.1$ in our implementation).

\subsection{Expert specialization}
The MoE framework comprises three anatomically specialized diffusion denoising experts $E_i$:

\begin{itemize}
    \item $E_1$: Specializes in denoising the centrum semiovale, enhancing smooth white matter regions with minimal textural variation.
    \item $E_2$: Focuses on the sulcal and gyral cortex, preserving high-frequency details within complex cortical folds.
    \item $E_3$: Targets the grey-white matter junction, ensuring sharp delineation at the cortical-subcortical interface.
\end{itemize}

Using the expert-specific Gaussian noise schedule, the latent representation $z_0$ is noised as follows:
\begin{equation}
\label{z}
z_{t(i)} = \sqrt{\bar{\alpha}_{t(i)}} z_0 + \sqrt{1 - \bar{\alpha}_{t(i)}} \epsilon_{(i)}, \quad \epsilon_{(i)} \sim \mathcal{N}(0, I)
\end{equation}

To ensure effective specialization within the MoE diffusion model, each expert $E_i$ is trained using a combination of a fundamental diffusion-based denoising loss and an expert-specific task loss tailored to distinct MRI restoration objectives. The training loss for each expert is defined as follows:

\begin{equation}
\mathcal{L}_{E_i} = \mathcal{L}_{\text{diffusion}(i)} + \mathcal{L}_{\text{task}(i)}
\end{equation}
where the diffusion loss $\mathcal{L}_{\text{diffusion}(i)}$ encourages accurate noise estimation during the diffusion process:

\begin{equation}
\mathcal{L}_{\text{diffusion}(i)} = \mathbb{E}_{z_{t(i)}, \epsilon_{(i)}} \left[ \| E_i(z_{t(i)}, t) - \epsilon_{(i)} \|_2^2 \right]
\end{equation}
where $z_{t(i)}$ is the noisy latent code at diffusion timestep $t$, and $\epsilon_{(i)}$ represents the Gaussian noise specifically introduced by expert $E_i$ during the forward diffusion process, while $E_i(z_{t(i)}, t)$ represents the predicted noise by each expert.

Following the diffusion framework described in Eq. \ref{z}, each expert estimates the clean latent representation $\hat{z}_{0(i)}$ through the reverse diffusion step:

\begin{equation}
\hat{z}_{0(i)} = \frac{z_{t(i)} - \sqrt{1 - \bar{\alpha}_{t(i)}} E_i(z_{t(i)}, t_i)}{\sqrt{\bar{\alpha}_{t(i)}}}
\end{equation}

The final latent representation $\hat{z}_0$ is computed by aggregating expert outputs weighted by the gating probabilities:

\begin{equation}
\hat{z}_0 = \sum_{i=1}^{K} \mathcal{G}_i \cdot \hat{z}_{0(i)}
\end{equation}
where $\mathcal{G}_i$ denotes the gating probability assigned to expert $E_i$, dynamically adjusting its contribution based on regional characteristics within the MRI slice. $K$ denotes the total number of experts in our approach.

The expert-specific task losses $\mathcal{L}_{\text{task}(i)}$ further refine expert specialization, addressing distinct anatomical or texture-related challenges:

\subsubsection{Expert 1}
To effectively denoise the centrum semiovale, an area dominated by smooth white matter structures with low textural variation, this expert emphasizes global structural coherence. It incorporates a dilated convolutional block in the bottleneck to enlarge the receptive field without increasing parameter count, capturing broad anatomical continuity. The encoder consists of an initial $3\times3$ convolutional layer, one residual block, and a lightweight Edge Enhancement Block utilizing a Laplacian kernel, followed by two downsampling convolutional layers ($4\times4$, stride 2) with residual blocks, totalling 8 layers. The bottleneck includes one $3\times3$ dilated convolution and one residual block (2 layers total). The decoder mirrors the encoder with two transposed convolutional layers and residual connections (8 layers), finalized by a $1\times1$ convolution. To promote perceptual fidelity of smooth structural regions, this expert incorporates a perceptual loss derived from a pre-trained VGG-16 \cite{simonyan2014very} feature extractor:

\begin{equation}
\mathcal{L}_{\text{task}(1)} = \| z_0 - \hat{z}_{0(1)} \|_2^2 + \|\Phi(z_0) - \Phi(\hat{z}_{0(1)})\|_2^2
\end{equation}
where $\Phi(\cdot)$ denotes the VGG-based perceptual feature extractor applied directly in latent space.

\subsubsection{Expert 2}
To enable more accurate extraction of cortical edges and high-frequency information, this expert adopts a complex convolutional bottleneck. The encoder consists of an initial $3\times3$ convolutional layer and two downsampling convolutional layers ($4\times4$, stride 2), each followed by a DenseBlock with four layers. In each DenseBlock, all preceding feature maps are concatenated as input to each subsequent layer, with a growth rate of 16 channels per layer. The bottleneck includes a Complex Convolution Block that jointly models magnitude and phase components. The decoder mirrors the encoder with two transposed convolutional layers and DenseBlocks, followed by a final $1\times1$ convolution to produce the output. For this, its loss function combining edge-aware and frequency-domain constraints is defined as:

\begin{equation}
\mathcal{L}_{\text{task}(2)} = \| S(z_0) - S(\hat{z}_{0(2)}) \|_2^2 + \| \mathcal{F}(z_0) - \mathcal{F}(\hat{z}_{0(2)}) \|_2^2
\end{equation}
where \( S(\cdot) \) is the Sobel operator and \( \mathcal{F}(\cdot) \) denotes the Fourier transform in latent space.

\subsubsection{Expert 3}
This expert is designed to delineate the grey–white matter junction, a region characterized by sharp structural transitions and subtle anatomical textures. To capture both long-range semantic dependencies and localized spectral variations, the architecture integrates a non-local attention block and a channel attention-enhanced decoding pathway. The encoder consists of an initial $3\times3$ convolutional layer, followed by two downsampling convolutional layers ($4\times4$, stride 2), each paired with a convolutional block and a channel attention module to strengthen region-specific contrast. The bottleneck includes one standard convolutional layer and one residual block to maintain compact semantic representation. The decoder mirrors the encoder with two transposed convolutional layers, channel attention blocks, and a Non-Local Attention Block that models spatial correlations across distant regions. The non-local module employs $1\times1$ convolutions to compute query, key, and value feature maps, followed by softmax-based attention aggregation and feature fusion. To enhance sensitivity to spatially localized frequency components, we employ a multi-scale Short-Time Fourier Transform (STFT) \cite{griffin1984signal}-based reconstruction loss:

\begin{equation}
\mathcal{L}_{\text{task}(3)} = \sum_{s \in \mathcal{S}} \sum_{w \in \Omega_s} \| \text{STFT}_s(z_0^w) - \text{STFT}_s(\hat{z}_{0(3)}^w) \|_2^2
\end{equation}
where \( \mathcal{S} \) denotes a predefined set of window sizes of \( \{8, 16, 32\} \), and \( \Omega_s \) is the set of sliding windows applied to the latent representation at scale \( s \). \( \text{STFT}_s(\cdot) \) denotes the STFT computed over each window.

\subsection{Joint training}
To facilitate effective region-adaptive learning in the MoE framework, we jointly train the expert networks and the gating mechanism through an integrated loss function. Specifically, the joint training incorporates both the expert-specific loss and a gating regularization term to promote accurate and robust expert selection. The overall expert loss is defined as a weighted sum of each expert's individual loss, dynamically modulated by gating probabilities:

\begin{equation}
\mathcal{L}_{E} = \sum_{i=1}^{K} \mathcal{G}_i \cdot \mathcal{L}_{E_i}
\end{equation}

To guide the gating network toward accurate expert assignments, we introduce a supervised expert-selection constraint based on the similarity between each expert’s predicted latent representation $\hat{z}_{0(i)}$ and the corresponding ground-truth latent representation $z_0$. To enhance numerical stability and robust estimation, we employ adaptive temperature scaling based on the Median Absolute Deviation (MAD) \cite{hampel1974influence}:

\begin{equation} 
T = 1.4826 \times \text{MAD}\left(\cos\theta(\hat{z}_{0(i)}, z_0)\right) 
\end{equation}
where $1.4826$ is a normalization constant aligning MAD to the standard deviation of a normal distribution. Using this adaptively scaled temperature $T$, the supervised expert-selection probabilities $\mathcal{G}_i^*$ are computed as follows:

\begin{equation} 
\mathcal{G}_i^* = \frac{\exp\left(\cos\theta(\hat{z}_{0(i)}, z_0) / T\right)}{\sum_{j=1}^{K} \exp\left(\cos\theta(\hat{z}_{0(j)}, z_0) / T\right)} 
\end{equation}

Then, to maintain expert specialization and avoid mode collapse, we incorporate a diversity-promoting regularization term into the gating loss. Instead of penalizing overall similarity across all expert pairs, we specifically target the most similar pairs, which ensures effective differentiation between experts while minimizing computational overhead:

\begin{equation}
\begin{aligned}
\mathcal{L}_{\text{gating}} =\ 
& \underbrace{-\frac{1}{K} \sum_{i=1}^{K} \mathcal{G}_i^* \log \mathcal{G}_i}_{\text{Supervised gating loss}} \\
& + \underbrace{\log\left(1 + \frac{1}{K} \sum_{i=1}^{K} \max_{j \neq i} \cos\theta(\hat{z}_{0(i)}, \hat{z}_{0(j)}) \right)}_{\text{Expert diversity regularization}}
\end{aligned}
\label{eq:gating_loss}
\end{equation}

Furthermore, to dynamically balance the relative contributions of the expert loss and gating loss, we apply an uncertainty-based adaptive weighting strategy, automatically adjusting the influence of $\mathcal{L}_{\text{gating}}$ based on its variance. This weighting term is defined as:

\begin{equation} 
w = \frac{1}{\sigma_{\text{gating}}^2} 
\end{equation}
where $\sigma_{\text{gating}}^2$ represents the variance of the gating loss, estimated through an EMA with a decay factor of 0.99.

Finally, the total joint training objective is formulated by combining the expert loss and adaptively weighted gating loss:

\begin{equation}
\label{final_loss}
\mathcal{L}_{\text{total}} = \mathcal{L}_{E} + w \mathcal{L}_{\text{gating}}
\end{equation}

\section{Experimental data and settings}

\subsection{Employed database}
The Human Connectome Project (HCP) \cite{van2013wu} maps the healthy human connectome by collecting neuroimaging and behavioural data on 1,200 normal young adults, aged 22-35. The project was carried out in two phases by a consortium of over 100 investigators and staff at 10 institutions. In Phase I, data acquisition and analysis methods were optimized, including refinements to pulse sequences and key preprocessing steps. In Phase II, neuroimaging and behavioural data were acquired from 1,200 healthy adults recruited from 300 families of twins and their non-twin siblings. To obtain brain connectivity maps of the highest quality, HCP employed cutting-edge MR hardware, including 3T and 7T MR scanners and customized head coils.

\subsection{Data preprocessing}
\label{data_preprocessing}
After normalization and standardization, each 3T and 7T MRI volume with a respective voxel size of 1.5mm $\times$ 1.5mm $\times$ 1.5mm and 0.7mm $\times$ 0.7mm $\times$ 0.7mm are resampled to a consistent size of $260 \times 310 \times 260$ voxels, where each slice is reformatted to a spatial resolution of $256 \times 256 \times 3$ pixels. The resulting dataset consists of 184 pairs of T1-weighted (T1w) MRI volumes at both 3T and 7T resolutions. These MRI slice pairs are then randomly divided into training, validation and test sets with a ratio of 8:1:1.

\subsection{Experimental details}
The VQ-VAE model was configured with a base channel size of 128, the embedding dimension was defined as 512, and the codebook size was 1024. During the MoEDiff-SR training, a base learning rate of 1e-06 was applied, with a lambda linear scheduler incorporating a warm-up period every 100 steps. The batch size was set to 32. The scale factor of the latent space was set as 0.2. The AdamW optimizer \cite{loshchilov2019} was utilized for training over 5,000 epochs. To ensure stable optimization of each expert at the early stage, the gating probabilities were fixed to a uniform distribution across all experts during the first 1,000 training epochs. For diffusion timesteps $T$, 500, 2000, and 1000 are configured for Expert 1, Expert 2, and Expert 3, respectively. We implemented our global approach using PyTorch v1.12.1 \cite{pytorch} on Nvidia A100 80 GB graphics cards.

\section{Experiments and results}
\subsection{Ablation study for incorporating bias filed and gradient nonlinearity correction}
\begin{figure}[htbp]
\centering
\includegraphics[width=0.45\textwidth]{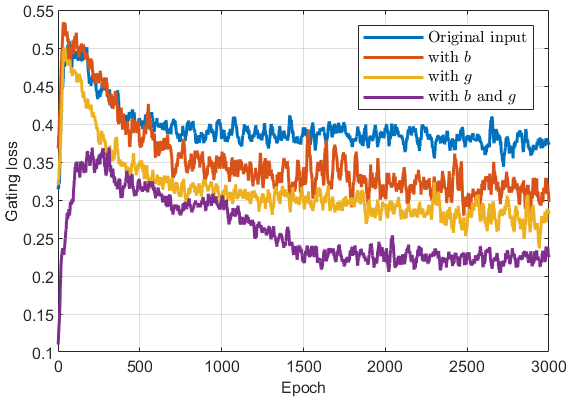}
\caption{Convergence analysis of gating loss under different configurations.}
\label{convergence_study}
\end{figure}

In our proposed MoE-based SR framework, each input 7T MRI slice is complemented by its corresponding bias field correction $b$ and gradient nonlinearity correction $g$. Such incorporation of these corrections enriches the contextual information available to the gating network, thereby enhancing its capacity for adaptive expert selection across diverse anatomical regions. To systematically evaluate the impact of these corrections, we conducted an ablation study, with the results presented in Fig. \ref{convergence_study}, illustrating the convergence behaviour of the gating loss across different configurations. As can be seen, the removal of either correction detrimentally affects training stability and convergence speed, leading to increased gating loss. In particular, the exclusion of both $b$ and $g$ results in the highest gating loss throughout training, indicating suboptimal expert specialization and heightened uncertainty in the gating mechanism. Among the individual corrections, gradient nonlinearity correction $g$ exhibits a more pronounced effect in reducing gating loss, suggesting its stronger influence in facilitating expert specialization, which could be attributed to the fundamental role of gradient nonlinearity in defining structural integrity within MRI images. Since tissue contrast and edge definition are largely influenced by gradient distortions, the absence of $g$ leads to greater ambiguity in anatomical boundaries, making it more difficult for the gating network to accurately assign expert contributions. In contrast, while the bias field correction $b$ plays an important role in intensity normalization, its effect is more global and less directly tied to fine-grained structural information, leading to a comparatively smaller impact on gating stability. Our proposed approach, incorporating both $b$ and $g$, consistently attains the lowest gating loss, signifying improved training stability and a more effective gating strategy. These findings underscore the critical role of bias field and gradient nonlinearity corrections in enhancing the robustness and precision of expert selection within the MoE framework, ultimately contributing to more accurate and stable SR performance.

Additionally, we present a quantitative evaluation of the SR performance on the test set using the Learned Perceptual Image Patch Similarity (LPIPS) metric \cite{zhang2018unreasonable}, as summarized in Table \ref{ablation_study}. The LPIPS metric quantifies the perceptual similarity between the generated 7T-like outputs and the corresponding 7T ground truth, with lower scores indicating higher perceptual fidelity and better alignment with the high-field MRI reference. As can be seen, incorporating both $g$ and $b$ yields the lowest LPIPS score of 0.0311. Removing either correction degrades performance, with the exclusion of $g$ leading to an increase in LPIPS to 0.0460, while omitting $b$ results in an even higher score of 0.0541. The absence of both corrections further exacerbates the perceptual degradation, yielding the highest LPIPS of 0.0598. For context, we report the LPIPS score of 0.0896 between real 3T and 7T images as a reference baseline, illustrating that even in the worst-case scenario without $g$ and $b$, our SR model outperforms the direct perceptual discrepancy between different field strengths. 

These results highlight the critical role of gradient nonlinearity and bias field corrections in enhancing both training stability and perceptual fidelity in our MoE-based super-resolution framework. The ablation study confirms that their absence significantly degrades performance, reinforcing their necessity for achieving more stable training and superior SR reconstruction quality.

\begin{table}[htbp]
\centering
\caption{Analysis of impact of $b$ and $g$}
\label{ablation_study}
\setlength{\tabcolsep}{4mm} 
\begin{tabular}{c|cc}
\toprule
\diagbox[width=18.6em]{$g$}{\scriptsize LPIPS}{$b$} 
& with $b$ & without $b$ \\ 
\hline
\\[-1.5ex] 
with $g$ & \bf 0.0311 & 0.0541 \\ 
without $g$ & 0.0460 & 0.0598 \\ 
\midrule
real 3T vs real 7T (context reference) & \multicolumn{2}{c}{0.0896}\\
\bottomrule
\end{tabular}
\end{table}

\subsection{Comparisons with state-of-the-art methods}
The comparison is divided into two main parts: Qualitative visualization analysis and quantitative metric-based analysis.

\subsubsection{Qualitative visualization analysis}
To facilitate a focused qualitative assessment, Table \ref{visualization} provides a visualization comparison of representative models from the CNN- and diffusion-based models, ESRGAN \cite{wang2018esrgan} and SR3 \cite{saharia2022image}, together with our proposed method. To ensure fairness and consistency, all models were trained on the same dataset using input slices with uniform dimensions (256 $\times$ 256 pixels). Given the voxel size transformation involved in the SR task, enhancing 2D in-plane resolution from 1.5mm$\times$1.5mm to 0.7mm$\times$0.7mm, an approximate magnification factor of 2.1$\times$ was required. To align with this target, ESRGAN was evaluated using its nearest supported scaling factor of 2$\times$, while SR3 was configured to produce outputs at 512$\times$512 pixels. For a more intuitive and comprehensive evaluation, regions clearly displaying distinctions among white matter, grey matter fissures, and cortical boundaries were specifically emphasized. Visualization results illustrate significant differences among these methods. As can be seen, ESRGAN tends to lose substantial high-frequency details, resulting in overly smooth outputs that lack crucial anatomical details and lead to overall image distortion. Conversely, SR3 demonstrates notably better SR performance on high-frequency edges, particularly cortical boundaries. However, SR3 still exhibits significant artefacts in cortical and white matter areas, undermining overall anatomical fidelity. Leveraging the MoE architecture, our proposed method effectively addresses the limitations observed in existing techniques. The collaboration among meticulously designed experts and different denoising strategies applied to each region enhance the simultaneous reconstruction capabilities for both high- and low-frequency regions, enabling accurate depiction of intricate cortical boundaries and coherent preservation of textures in grey and white matter areas, closely resembling the original 7T ground-truth MRIs.

\begin{table*}[htbp]
\centering
\caption{Qualitative visualization comparison with SOTA methods}
\label{visualization}
\begin{threeparttable}
\setlength{\tabcolsep}{0.2mm}
\begin{tabular}{cccccc}
\toprule
 Input (3T) & Ground Truth (7T)& ESRGAN \cite{wang2018esrgan} & SR3 \cite{saharia2022image} & Ours\\
\midrule
\begin{minipage}[b]{0.4\columnwidth}\centering \raisebox{-.5\height}{\includegraphics[width=\linewidth]{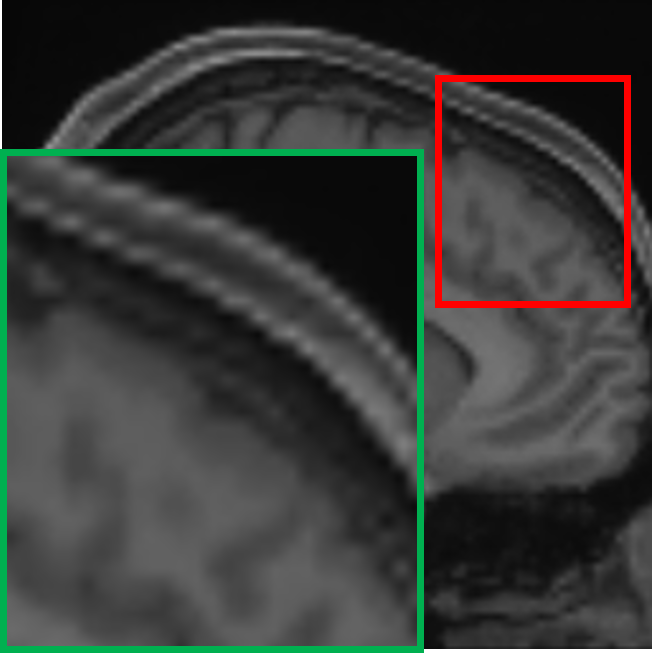}}\end{minipage}&\begin{minipage}[b]{0.4\columnwidth}\centering \raisebox{-.5\height}{\includegraphics[width=\linewidth]{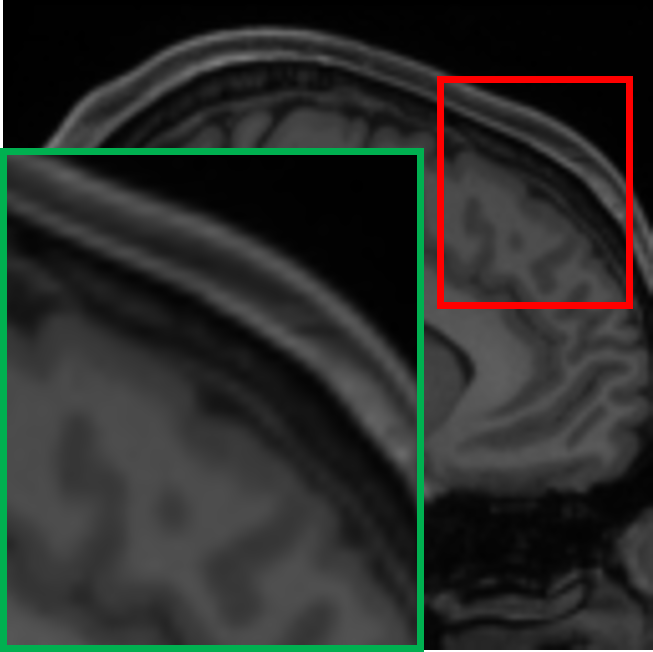}}\end{minipage}&\begin{minipage}[b]{0.4\columnwidth}\centering \raisebox{-.5\height}{\includegraphics[width=\linewidth]{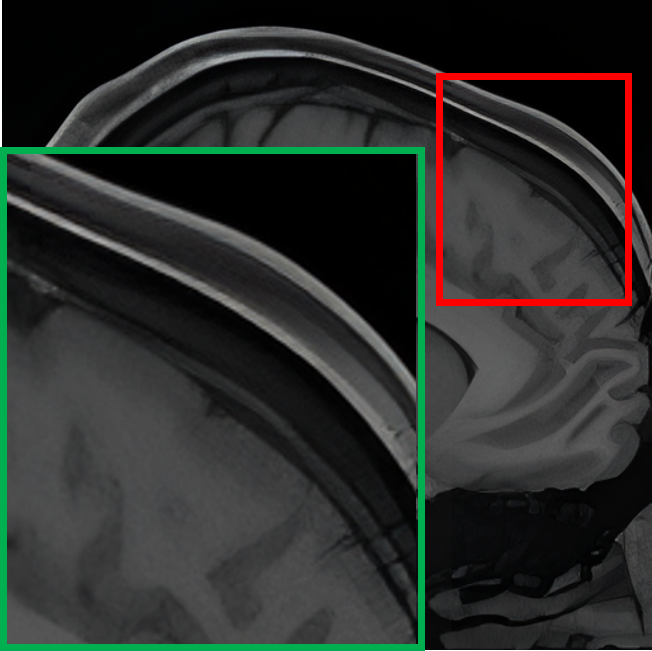}}\end{minipage}&\begin{minipage}[b]{0.4\columnwidth}\centering \raisebox{-.5\height}{\includegraphics[width=\linewidth]{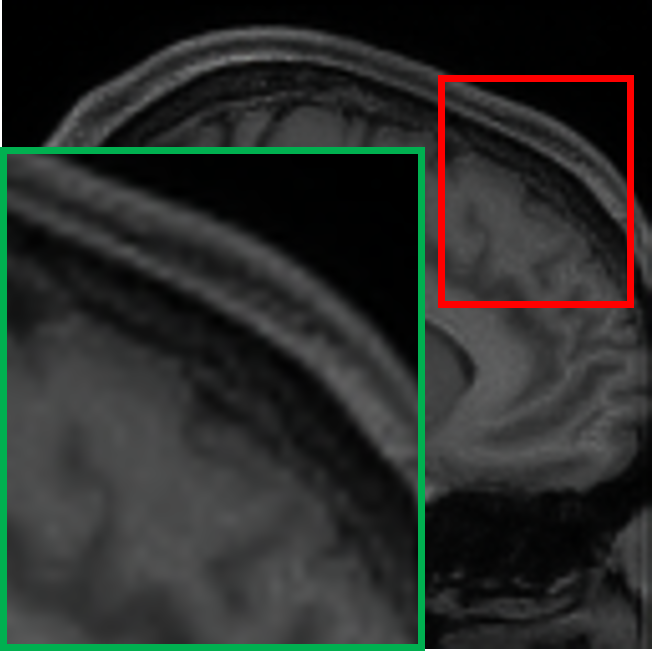}}\end{minipage}&\begin{minipage}[b]{0.4\columnwidth}\centering \raisebox{-.5\height}{\includegraphics[width=\linewidth]{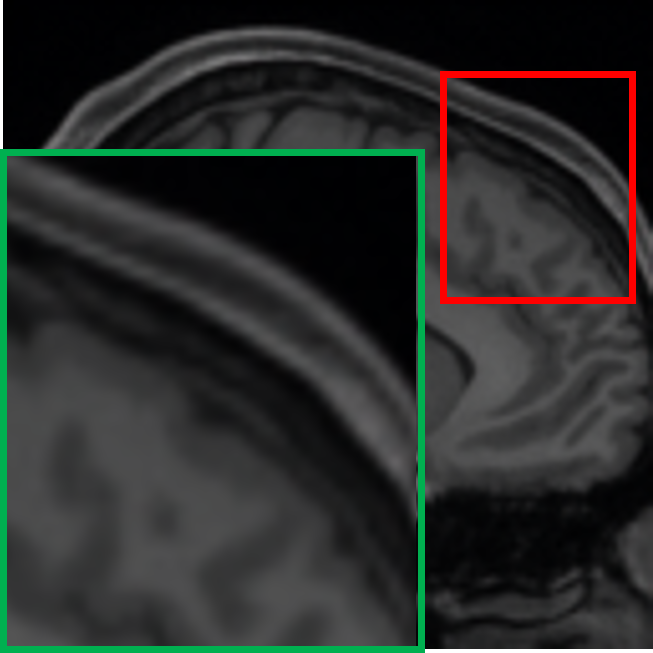}}\end{minipage}\\
\bottomrule
\end{tabular}
\end{threeparttable}
\end{table*}

\subsubsection{Quantitative metric-based analysis}
Beyond ESRGAN and SR3, Table \ref{metrics} extends the comparison to a diverse set of State-of-the-Art (SOTA) SR methods, including WavTrans \cite{li2022wavtrans}, MASA \cite{lu2021masa}, McMRSR \cite{li2022transformer}, MC-VarNet \cite{lei2023decomposition}, and DisC-Diff \cite{mao2023disc}, thereby enabling a comprehensive performance evaluation. The assessment covers image quality metrics in terms of PSNR (dB), SSIM, LPIPS ($10^{-2}$), and RMSE ($10^{-2}$), as well as computational efficiency, measured in terms of Params ($M$) and FLOPs ($G$). As can be seen, all selected methods have achieved varying degrees of SR performance based on the 3T data. However, our proposed approach consistently achieves the best overall performance across all image quality metrics. Specifically, it attains the highest PSNR of 38.16 dB and SSIM of 0.93, indicating superior fidelity and structural preservation, while also achieving the lowest LPIPS of 1.18 and RMSE of 1.83, which reflect enhanced perceptual quality and minimal reconstruction error. These results underscore the model’s ability to generate highly accurate and visually faithful super-resolved images.  Among the competing diffusion-based methods, SR3 and DisC-Diff, although they trail our method across all metrics, nonetheless demonstrate the competitive efficacy of advanced diffusion architectures. MC-VarNet, which is based on a variational network framework, achieves reconstruction quality (PSNR of 35.59, SSIM of 0.89), but exhibits inferior perceptual alignment (LPIPS of 2.12) and higher reconstruction error (RMSE of 3.47), suggesting limitations in modelling finer image textures. Intermediate performance is observed in MASA and WavTrans, highlighting their limited capacity in recovering both global structure and fine-scale details. From a computational perspective, diffusion-based models are typically characterized by relatively large parameter counts, attributable to their inherently complex generative architectures. Nevertheless, they exhibit notable advantages in computational efficiency in terms of FLOPs. For example, although our ensemble diffusion framework comprises a substantial number of parameters (133.02 $M$), it maintains a low computational cost of only 56.12 $G$ FLOPs, primarily due to its latent-space operation. This stands in sharp contrast to CNN-based models such as ESRGAN, which, despite having a comparatively modest parameter count (16.73 $M$), incurs a substantially higher computational burden of 1450.87 $G$ FLOPs. This inefficiency arises largely from its reliance on pixel-level inference and densely connected convolutional operations at full spatial resolution, which collectively contribute to inflated computational demands. Furthermore, benefiting from the decoupled structure of the MoE architecture, which allows asynchronous activation of expert networks during inference, our model can further reduce the computational load to 42.21 $G$ FLOPs, engaging 62.69 $M$ parameters, which facilitates efficient adaptation to resource-constrained environments, without compromising model performance.

\begin{table}[htbp]
\centering
\caption{Quantitative comparison of image quality metrics and computational cost across methods}
\label{metrics}
\begin{threeparttable}
\setlength{\tabcolsep}{1.6mm} 
\begin{tabular}{lcccc|cc}
\toprule
& PSNR & SSIM & LPIPS & RSME  & Params & FLOPs \\ 
\hline
3T (ref) & 24.13 & 0.78 & 8.96 & 6.22& -& -\\
\hline
ESRGAN \cite{wang2018esrgan} & 24.45 & 0.82&10.53&6.59&16.73&1450.87\\
SR3 \cite{saharia2022image} & 34.49 & 0.89&3.79&2.49&97.81&46.32\\
WavTrans \cite{li2022wavtrans}& 31.07 & 0.85 & 5.39 & 5.17&\bf 2.1&162.89\\
MASA \cite{lu2021masa}& 33.70 & 0.88 &3.11&5.48&4.0&180.13\\
McMRSR \cite{li2022transformer}& 30.91 & 0.83 &6.07&5.73&3.5&269.86\\
MC-VarNet \cite{lei2023decomposition}& 35.59 & 0.89 &2.12&3.47&5.7&139.86\\
DisC-Diff \cite{mao2023disc}& 36.92 & 0.91 &1.67&2.13&86.1 & 461.00 \\
Ours  & \bf 38.16& \bf 0.93& \bf 1.18& \bf 1.83&133.02 & 56.12 \\
Ours (asyn)$^*$ &- &-&-&-&62.69  & \bf 42.21 \\
\bottomrule
\end{tabular}
\begin{tablenotes}
\footnotesize
\item[$*$] The reported results are from the most computationally intensive expert network during asynchronous inference.
\end{tablenotes}
\end{threeparttable}
\end{table}

Overall, our proposed approach achieves SOTA image quality across multiple quantitative metrics while maintaining high computational efficiency. These results underscore its effectiveness in balancing reconstruction fidelity with resource-efficient deployment.

\subsection{Visualization of the expert specialization}
To further elucidate the functional specialization of each expert within the proposed MoE framework, we visualize the difference maps between each expert’s output and the ground truth 7T MRI. These maps are computed by taking the pixel-wise absolute difference between the expert output and the ground truth 7T MRI, highlighting the residual discrepancies and reconstruction focus of each expert. In the visualizations, lighter areas reflect lower reconstruction errors, suggesting better alignment with the ground truth. As shown in Fig. \ref{expert_visualization}, each column represents the output of a particular expert, visualized through its corresponding absolute difference map with respect to the 7T reference. The final output is computed via a weighted combination of expert outputs. The visualized difference maps underscore the distinct specialization of each expert. Specifically, Expert 1 exhibits minimal residuals in homogeneous white matter regions, preserving large-scale anatomical continuity. Its low-frequency focus is evident in the smooth residual patterns and coherent structural boundaries. Expert 2 demonstrates heightened sensitivity to cortical regions and sulcal boundaries, with pronounced residual suppression around edges and high-frequency details. Expert 3 shows enhanced reconstruction accuracy at tissue interfaces and transitional zones. The residuals are particularly reduced in regions of sharp contrast changes. These observations confirm that each expert effectively targets specific anatomical and textural characteristics, guided by their unique architectural and loss function design. The gating mechanism further facilitates optimal integration by adaptively weighting expert contributions according to the local image context, thus ensuring a globally coherent and anatomically faithful 7T-like reconstruction.

\begin{table}[htbp]
\centering
\caption{Visualization of the expert specialization}
\label{expert_visualization}
\setlength{\tabcolsep}{0.3mm}
\begin{threeparttable}
\begin{tabular}{cccc}
\toprule
$E_1$ & $E_2$ & $E_3$& Final\\
\midrule
\begin{minipage}[b]{0.24\columnwidth}\centering \raisebox{-.5\height}{\includegraphics[width=\linewidth]{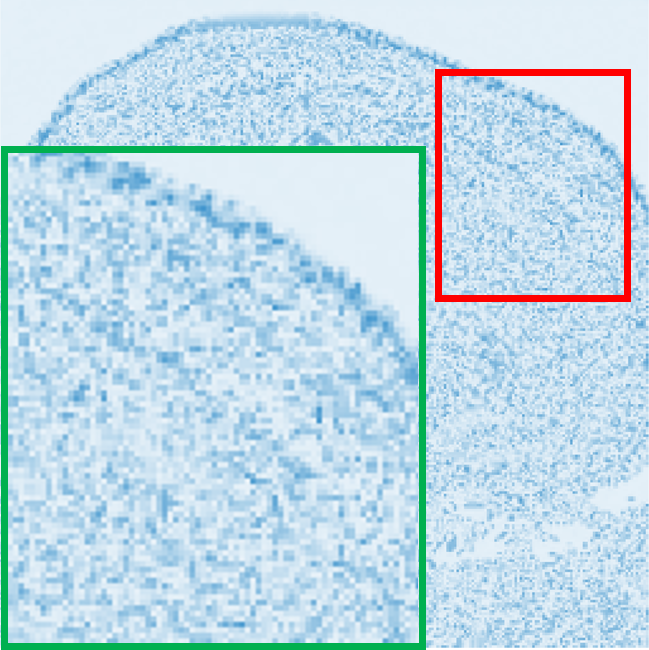}}\end{minipage}&\begin{minipage}[b]{0.24\columnwidth}\centering \raisebox{-.5\height}{\includegraphics[width=\linewidth]{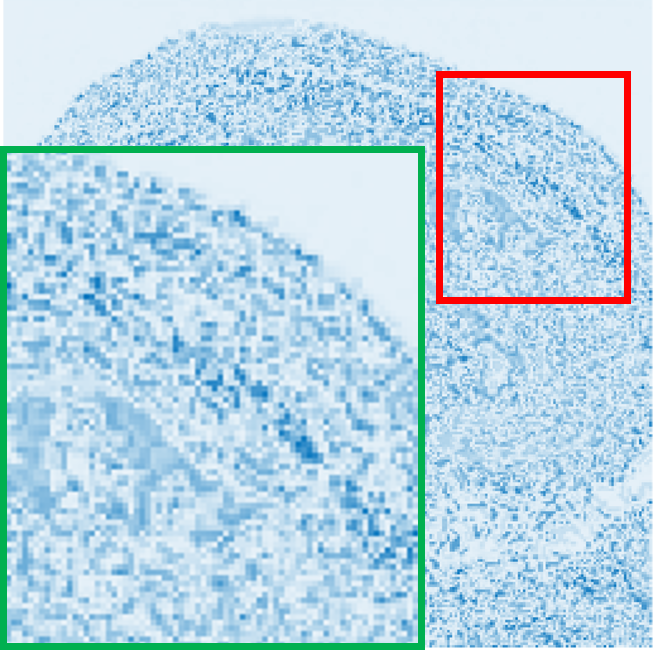}}\end{minipage}&\begin{minipage}[b]{0.24\columnwidth}\centering \raisebox{-.5\height}{\includegraphics[width=\linewidth]{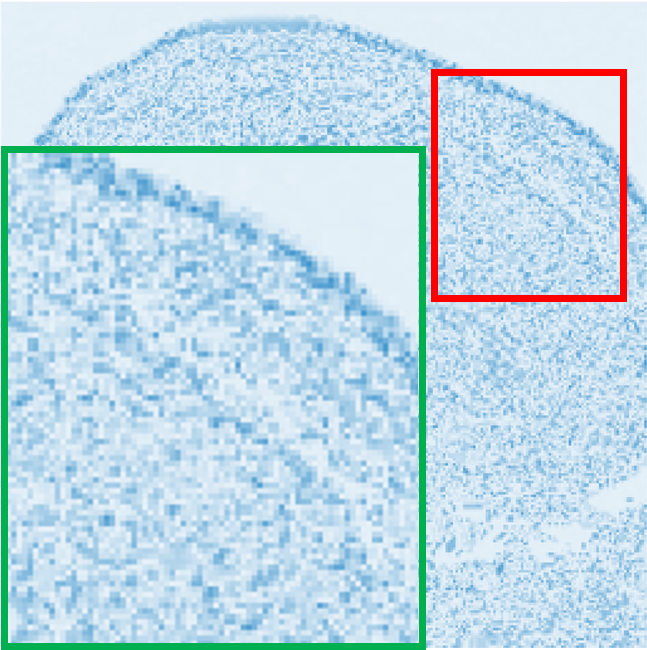}}\end{minipage}&\begin{minipage}[b]{0.24\columnwidth}\centering \raisebox{-.5\height}{\includegraphics[width=\linewidth]{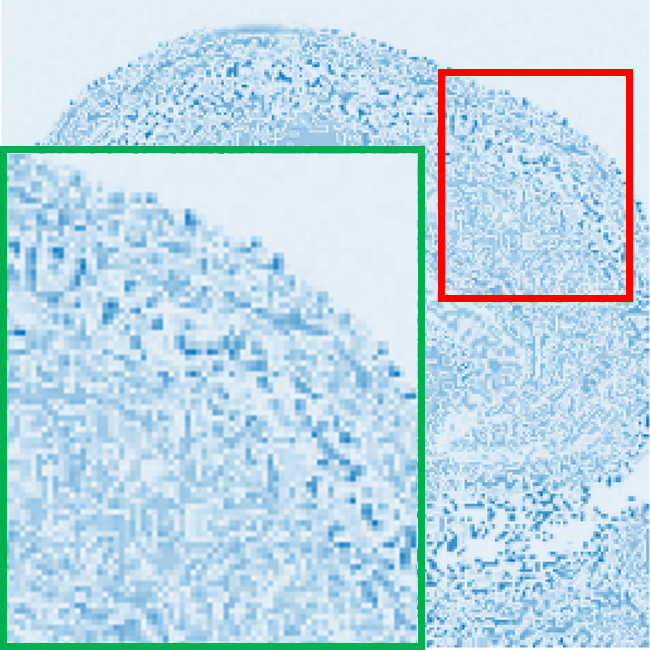}}\end{minipage}\\
\midrule
$\mathcal{G}_1=0.21$ & $\mathcal{G}_2=0.48$  & $\mathcal{G}_3=0.31$  & Weighted\\
\bottomrule
\end{tabular}
\begin{tablenotes}
\footnotesize
\item[$*$] The expert-specific weights ($\mathcal{G}_1$, $\mathcal{G}_2$, $\mathcal{G}_3$) for this particular slice are dynamically computed by the gating network $\mathcal{G}$.
\end{tablenotes}
\end{threeparttable}
\end{table}

\subsection{Clinical evaluation}
To ensure the clinical validity of our proposed SR framework, ethical approval was obtained from the Institutional Review Board (IRB) at Massachusetts General Hospital (MGH) under protocol number 2024P003489. All datasets used in this study were fully de-identified to adhere to strict ethical guidelines and privacy regulations.

\begin{table}[htbp]
\centering
\caption{Clinical evaluation for real cases at MGH}
\label{clinical_2}
\setlength{\tabcolsep}{0.1mm}
\begin{tabular}{ccc}
\toprule
Real 3T MRI &  Real 7T MRI &  7T-like SR MRI\\
 \midrule
\begin{minipage}[b]{0.33\columnwidth}\centering \raisebox{-.5\height}{\includegraphics[width=\linewidth]{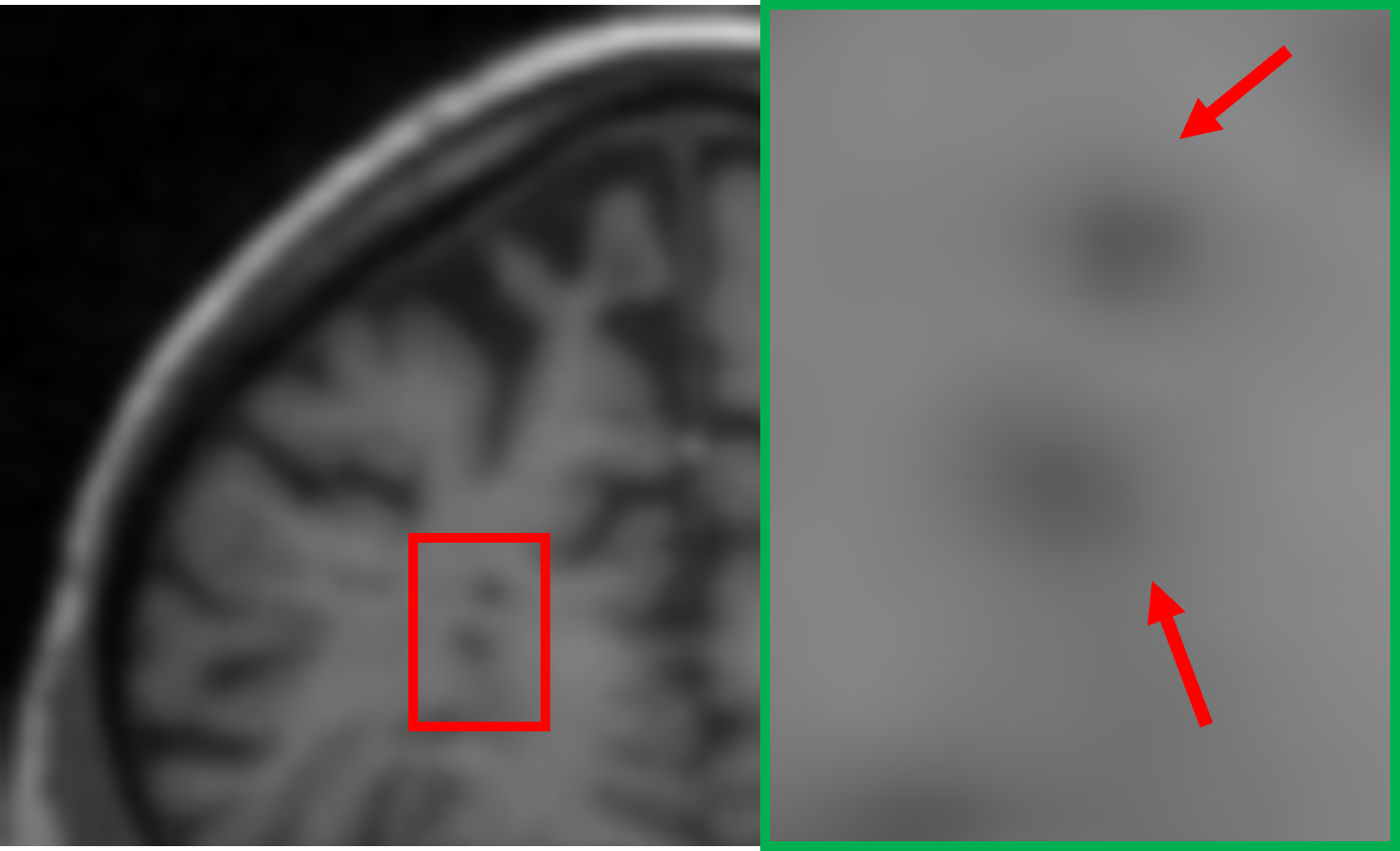}}\end{minipage}&\begin{minipage}[b]{0.33\columnwidth}\centering \raisebox{-.5\height}{\includegraphics[width=\linewidth]{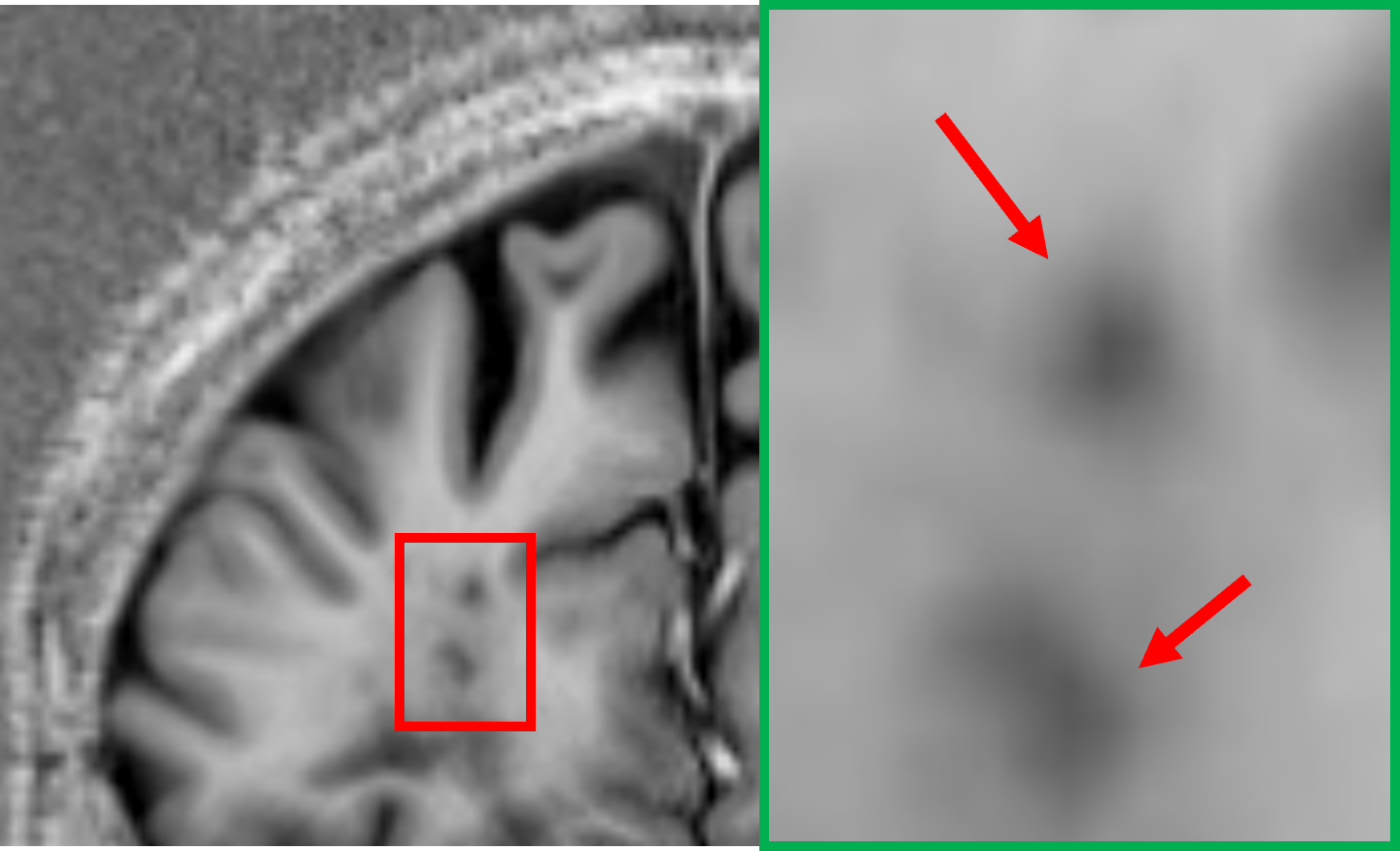}}\end{minipage}&\begin{minipage}[b]{0.33\columnwidth}\centering \raisebox{-.5\height}{\includegraphics[width=\linewidth]{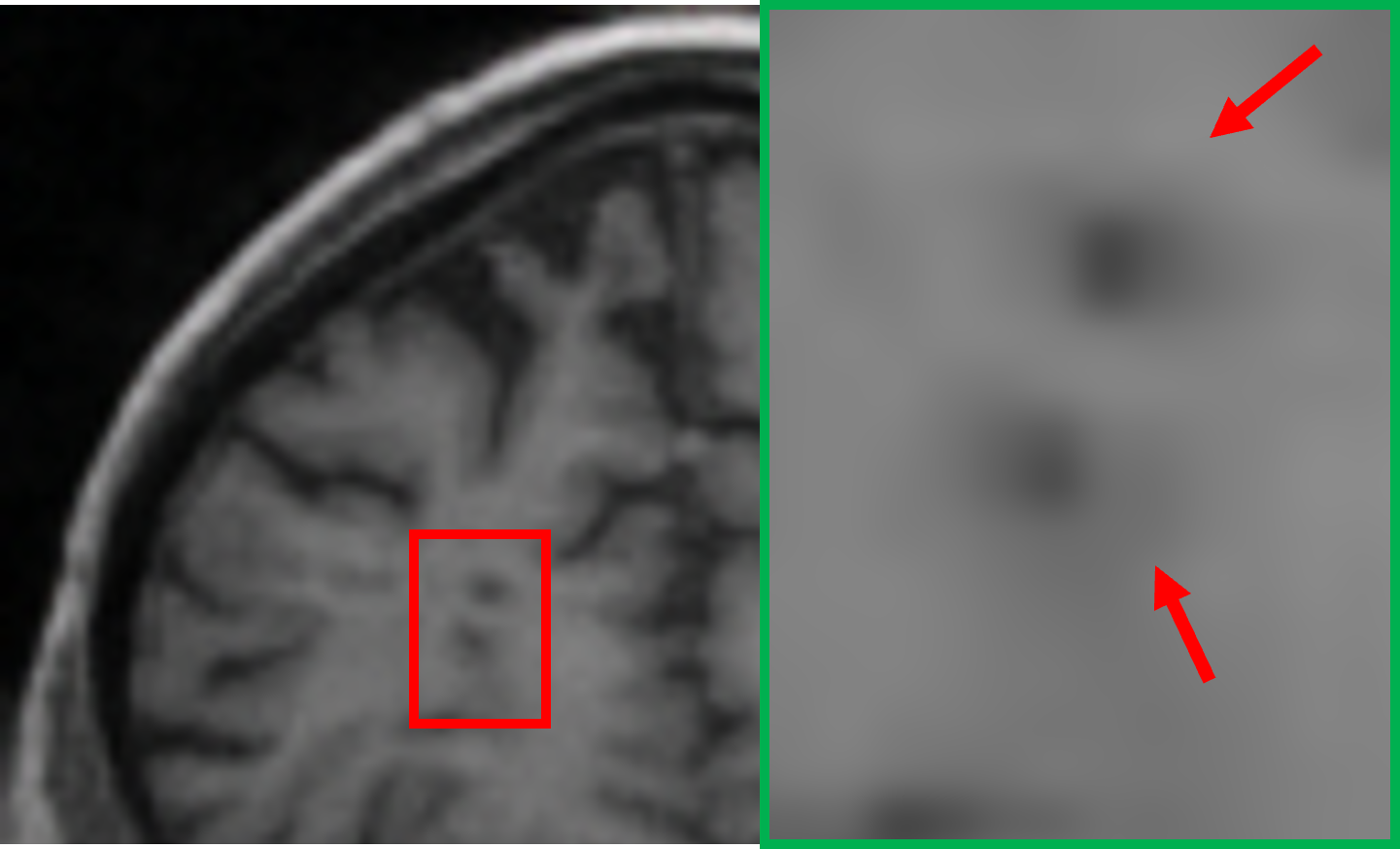}}\end{minipage}\\
\bottomrule
\end{tabular}
\end{table}

In a representative clinical case of Multiple Sclerosis (MS), a demyelinating disease characterized by periventricular and subcortical white matter abnormalities, lesion detection poses a significant challenge, particularly on lower-field MRI systems. MS lesions, often presenting as small hypointense foci, so-called “black holes”, on T1w imaging, are frequently subtle and prone to under-recognition at 3T due to limitations in spatial resolution and tissue contrast. As illustrated in Table \ref{clinical_2}, 7T MRI provided a marked improvement in image resolution and contrast, enabling enhanced visualization of periventricular abnormalities and facilitating the detection of MS-associated pathology with greater fidelity, which highlights the diagnostic advantages of ultra-high-field imaging in neuroinflammatory conditions. It is noteworthy that, when our proposed SR method was applied to the 3T MRI, the synthesized 7T-like image closely approximated the ground-truth 7T scan in its delineation of periventricular white matter and the visibility of small hypointense lesions. While exact slice correspondence between 3T and 7T scans cannot be guaranteed due to inherent inter-scan variability in positioning and acquisition parameters, the qualitative improvements achieved by the SR model are evident in the enhanced conspicuity of clinically relevant features, which underscores the clinical utility of our proposed SR framework.

\subsection{Discussion}
In this study, we introduced MoEDiff-SR, a novel Mixture-of-Experts (MoE) guided diffusion framework for region-adaptive super-resolution (SR) of brain MRI. By integrating a Transformer-based gating mechanism with anatomically specialized diffusion experts, the proposed model adaptively modulates the denoising process in accordance with the structural and textural heterogeneity across brain regions. This expert-driven design enables precise enhancement of both low-frequency anatomical continuity and high-frequency cortical details, yielding superior performance over state-of-the-art CNN- and diffusion-based SR methods across perceptual, structural, and quantitative evaluation metrics. Clinical assessments further validate the diagnostic utility of the generated 7T-like images, particularly in scenarios where conventional 3T MRI fails to capture subtle but clinically relevant abnormalities. Overall, our proposed MoEDiff-SR effectively bridges data-driven SR with anatomy-aware modelling, offering a scalable and clinically meaningful solution for enhancing image quality in routine neuroimaging.
\subsubsection{Strengths}
A central innovation of MoEDiff-SR lies in its anatomically-informed denoising strategy. From a technical perspective, unlike prior SR methods that rely on globally uniform processing, our model adaptively distributes denoising responsibilities across specialized diffusion experts. As shown in both quantitative metrics and visual analysis, this architecture results in superior reconstruction fidelity, especially in areas where classical methods often struggle. Second, the integration of bias field correction and gradient nonlinearity correction into the latent embedding space further enhances the model’s capacity for accurate expert assignment, which suggests that domain-specific prior information can be repurposed as conditioning cues within learning-based frameworks to drive more anatomically-aware restoration. Another critical strength is the model’s deployment efficiency. Despite comprising multiple diffusion experts, the MoE design supports asynchronous inference, allowing individual experts to be selectively activated based on region-specific needs, which enables a substantial reduction in FLOPs and parameter usage without compromising image quality. From a clinical perspective, clinical evaluation has affirmed that MoEDiff-SR-generated 7T-like images offer not only improved visual quality but also enhanced diagnostic clarity, especially in cases where subtle pathological features are otherwise obscured at lower field strengths.

\subsubsection{Limitations}
Despite these strengths, the study has several limitations. Firstly, the performance of the model is heavily dependent on the availability of large, high-quality paired data. The scarcity of datasets, especially for rare conditions or specialized applications, may restrict the generalizability of the approach. Secondly, although the MoE strategy results in a significantly lighter inference process, the model still requires approximately 45GB of graphic memory, deploying the framework on common computers thus remains challenging. Finally, the use of bias field correction and gradient nonlinearity correction, while beneficial for generating high-quality outputs, introduces an additional dependency on pre-processing steps, which could pose challenges in workflows where these corrections are unavailable or impractical to implement consistently.

\subsubsection{Future work}
Future work for this study will focus on expanding the generalizability and robustness of the proposed SR framework. One key direction is the exploration of multi-modal data integration, combining MRI with complementary imaging modalities such as CT or ultrasound, to improve the reconstruction process and accuracy. Additionally, expanding the framework to accommodate other imaging modalities could further broaden its applicability and impact across diverse clinical and research domains.

\section{Statements}
This manuscript was prepared using data from the HCP project. The views expressed in it are those of the authors and do not necessarily reflect the opinions of the HCP investigators, the National Institutes of Health (NIH), or the private funding partners.

\section{Acknowledgements}
The authors gratefully acknowledge the support of the Ralph Schlaeger Research Fellowship under award number 246448 from Massachusetts General Hospital (MGH), Harvard Medical School (HMS) and the French National Agency of Research (ANR) under project number ANR-20-CE45-0013-01.

\bibliographystyle{IEEEtran}
\bibliography{cas-refs}

\end{document}